\documentclass[12pt]{paper}

\usepackage{epsfig}

\begin{document}

\centerline{\bf \Large Grand Unified Theories} \baselineskip=16pt
\vspace{0.8cm} \centerline{\rm \large Stuart Raby}
\baselineskip=14pt \centerline{\it Department of Physics, The Ohio
State University, 191 W. Woodruff Ave.} \baselineskip=12pt
\centerline{\it Columbus, OH 43210} \vspace{0.9cm}
\rm\baselineskip=13pt  \centerline{\it Invited talk given at the 2nd
World Summit on Physics Beyond the Standard Model}
\baselineskip=12pt \centerline{\it Galapagos Islands, Ecuador June
22-25, 2006}

\section{Grand Unification}

The standard model is specified by the local gauge symmetry group,
$SU(3) \times SU(2) \times U(1)_Y$, and the charges of the matter
particles under the symmetry.  One family of quarks and leptons
[${Q, u^c, d^c ; L, e^c}$] transform as [$(3,2,1/3),$ $({\bar 3},1,
-4/3),$ $({\bar 3},1, 2/3) ;$ $(1,2,-1),$ $(1,1,2)$], where $Q = (u,
d)$ and $L = (\nu, e)$ are $SU(2)_L$ doublets and $u^c,\; d^c, \;
e^c$ are charge conjugate $SU(2)_L$ singlet fields with the $U(1)_Y$
quantum numbers given. [We use the convention that electric charge
$Q_{EM} = T_{3 L} + Y/2$ and all fields are left handed.]   Quark,
lepton,  W and Z masses are determined by dimensionless couplings to
the Higgs boson and the neutral Higgs vev.    The apparent hierarchy
of fermion masses and mixing angles is a complete mystery in the
standard model.   In addition, light neutrino masses are presumably
due to a See-Saw mechanism with respect to a new large scale of
nature.

The first unification assumption was made by Pati and Salam [PS]
\cite{ps} who proposed that lepton number was the fourth color,
thereby enlarging the color group $SU(3)$ to $SU(4)$.   They showed
that one family of quarks and leptons could reside in two
irreducible representations of a left-right symmetric group  $SU(4)
\times SU(2)_L \times SU(2)_R$ with weak hypercharge given by $Y/2 =
1/2 (B - L) + T_{3 R}$.\footnote{Of course, the left-right symmetry
required the introduction of a right-handed neutrino contained in
the left handed Weyl spinor, $\nu^c$.} Thus in PS, electric charge
is quantized.

Shortly after PS was discovered, it was realized that the group
$SO(10)$ contained PS as a subgroup and unified one family of quarks
and leptons into one irreducible representation, {\bf 16}
\cite{SO(10)}. See Table \ref{t:table}.    This is clearly a
beautiful symmetry, but is it realized in nature?   If yes, what
evidence do we have?

\protect
\begin{table}
\caption{The quantum numbers of the ${\bf 16}$ dimensional representation
of $SO(10)$.}
\label{t:table}
$$\begin{array}{|c|c|c|c|}
\hline
{\rm State}  &  Y    &  {\rm Color} & {\rm Weak}  \\
\hline
\hline
{\bf  \nu^c}  &  0    &  + \, + \, + &     + \, +  \\
\hline
\hline
{\bf  e^c}  & 2   & + \, + \, +  &    - \, -  \\
\hline
{\bf u_r} &  1/3  &  - \, + \, +  &    + \, -   \\

{\bf d_r}  & 1/3  &  - \, + \, +  &   - \, +   \\
\hline
{\bf u_b}  &  1/3 &  + \, - \, +   &   + \, -   \\

{\bf d_b}  &  1/3 &  + \, - \, +  &   - \, +   \\
\hline
{\bf u_y}  &   1/3 &   + \, + \, -  &    + \, -  \\
{\bf d_y}  &  1/3  & + \, + \, -  & - \, +  \\
\hline
{\bf u_r^c}  &  -4/3 & + \, - \, -  & + \, +  \\

{\bf u_b^c}  & -4/3 & - \, + \, -  & + \, +   \\

{\bf u_y^c}  & -4/3 & - \, - \, +  & + \, +  \\
\hline
\hline
{\bf d_r^c}  & 2/3 & + \, - \, -  & - \, -   \\

{\bf  d_b^c}  &  2/3  & - \, + \, -  &  - \, -   \\

{\bf d_y^c}  &  2/3 & - \, - \, +  & - \, -  \\
\hline
{\bf \nu}  &  -1 & - \, - \, -  & + \, -  \\

{\bf e}  &  -1  &  - \, - \, -  &  - \, + \\
\hline
\end{array}$$
\end{table}

\begin{figure}[t!]
\begin{center}
\begin{minipage}{6.5in}
\epsfig{file=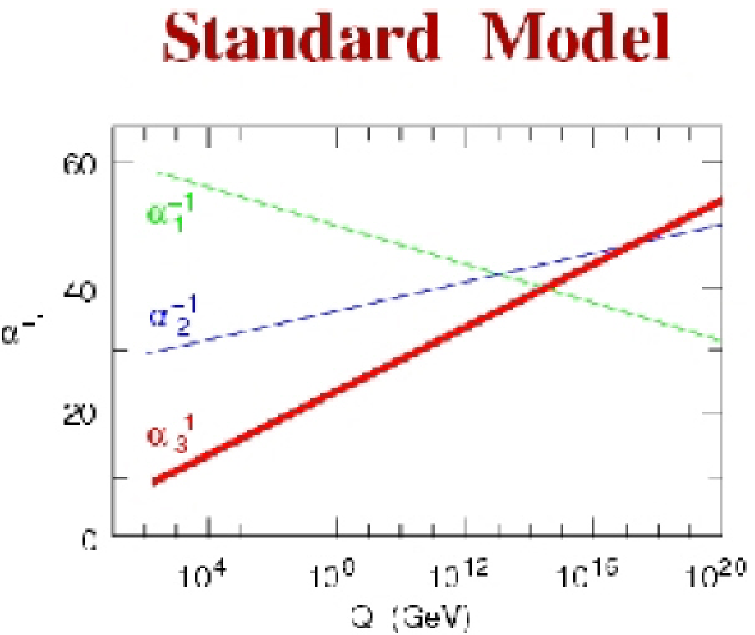,height=2.2in} \hspace{.3in}
\epsfig{file=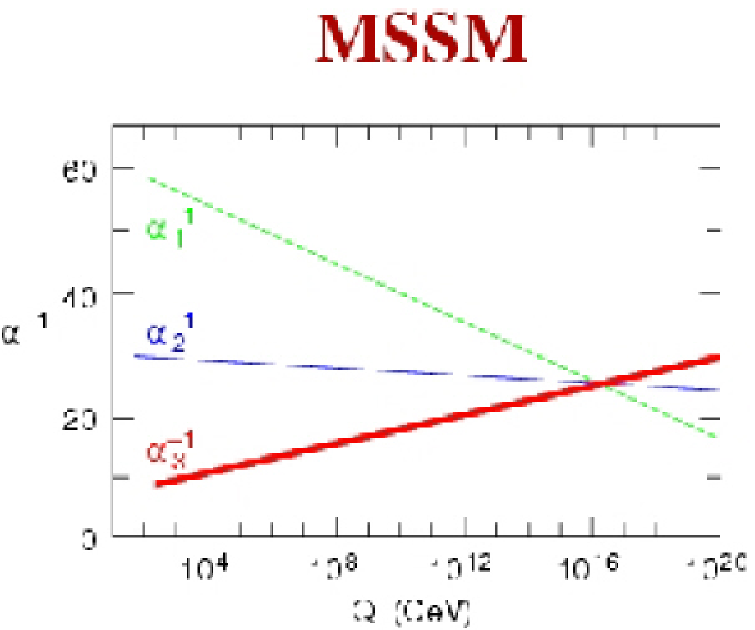,height=2.2in}
\end{minipage}
\caption{\label{fig:guts} {\small Gauge coupling unification in
non-SUSY GUTs (top) vs. SUSY GUTs (bottom) using the LEP data as of
1991. Note, the difference in the running for SUSY is the inclusion
of supersymmetric partners of standard model particles at scales of
order a TeV.}}
\end{center}
\end{figure}

Of course,  grand unification makes several predictions.   The first
two being  gauge coupling unification and proton decay
\cite{gg,gqw}.   Shortly afterward it was realized that Yukawa
unification was also predicted \cite{chanowitz}.   Experiments
looking for proton decay were begun in the early 80s.   By the late
80s it was realized that grand unification apparently was not
realized in nature (assuming the standard model particle spectrum).
Proton decay was not observed and in 1992,  LEP data measuring the
three standard model fine structure constants $\alpha_i, \ i =
1,2,3$ showed that non-supersymmetric grand unification was excluded
by the data \cite{adf}.    On the other hand,  supersymmetric GUTs
\cite{drw} (requiring superpartners for all standard model particles
with mass of order the weak scale) was consistent with the LEP data
\cite{adf} and at the same time raised the GUT scale; thus
suppressing proton decay from gauge boson exchange \cite{drw}. See
Fig. \ref{fig:guts}.

The spectrum of the minimal SUSY theory includes all the SM states
plus their supersymmetric partners.   It also includes one pair (or
more) of Higgs doublets; one to give mass to up-type quarks and the
other to down-type quarks and charged leptons.   Two doublets with
opposite hypercharge $Y$ are also needed to cancel triangle
anomalies.   Finally, it is important to recognize that a low energy
SUSY breaking scale (the scale at which the SUSY partners of SM
particles obtain mass) is necessary to solve the gauge hierarchy
problem.

SUSY extensions of the SM have the property that their effects
decouple as the effective SUSY breaking scale is increased. Any
theory beyond the SM must have this property simply because the SM
works so well.  However, the SUSY breaking scale cannot be increased
with impunity, since this would reintroduce a gauge hierarchy
problem.   Unfortunately there is no clear-cut answer to the
question, when is the SUSY breaking scale too high. A conservative
bound would suggest that the third generation quarks and leptons
must be lighter than about 1 TeV, in order that the one loop
corrections to the Higgs mass from Yukawa interactions remains of
order the Higgs mass bound itself.
\medskip

SUSY GUTs can naturally address all of the following issues:
\begin{itemize}
\item    $M_Z \ll M_{GUT}$   "Natural"
\item    Explains  charge  quantization
\item    Predicts  gauge coupling  unification  !!
\item     Predicts  SUSY particles at LHC
 \item    Predicts  proton  decay
 \item    Predicts  Yukawa coupling unification
\item     and with broken Family symmetry explains the fermion mass  hierarchy
\item     Neutrino masses and mixing via See-Saw
 \item    LSP -  Dark Matter candidate
 \item    Baryogenesis  via  leptogenesis
\end{itemize}

In this talk, I will consider two topics.  The first topic is a
comparison of GUT predictions in the original 4 dimensional field
theory version versus its manifestation in 5 or 6 dimensional
orbifold field theories (so-called orbifold GUTs) or in the 10
dimensional heterotic string.     The second topic focuses on the
minimal SO(10) SUSY model [MSO$_{10}$SM].   By definition, in this
theory the electroweak Higgs of the MSSM is contained in a single
{\bf 10} dimensional representation of SO(10).   There are many
experimentally verifiable consequences of this theory.   This is due
to the fact that \begin{enumerate} \item in order to fit the top,
bottom and tau masses one finds that the SUSY breaking soft terms
are severely constrained,  and  \item  the ratio of the two Higgs
vevs,  $\tan\beta \sim 50$.  This by itelf leads to several
interesting consequences.  \end{enumerate}

\section{Gauge coupling unification \& Proton decay \label{sec:guts}}

\subsection{4D GUTs}

At $M_G$ the GUT field theory matches on to the minimal SUSY low
energy theory with matching conditions given by $g_3 = g_2 = g_1
\equiv g_G$,  where at any scale $\mu < M_G$ we have $g_2 \equiv g$
and $g_1 = \sqrt{5/3} \; g^\prime$. Then using two low energy
couplings, such as $\alpha_{EM}(M_Z),\; \sin^2\theta_W$, the two
independent parameters $\alpha_G, \; M_G$ can be fixed.  The third
gauge coupling, $\alpha_s(M_Z)$ is predicted.

\begin{figure}[t!]
\hspace{-.70in}
\begin{center}
\begin{minipage}{5in}
\epsfig{file=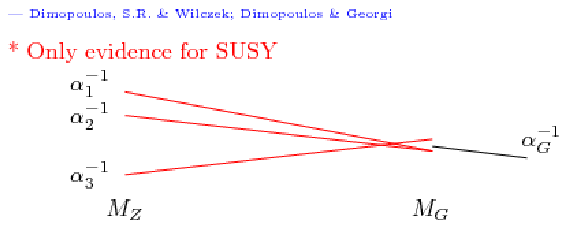,height=2in}
\end{minipage}
\caption{\label{fig:guts.2loop} {\small Gauge coupling unification
in SUSY GUTs using the LEP data. Given the present accurate
measurements of the three low energy couplings, in particular
$\alpha_s(M_Z)$, and the fact that the theoretical analysis now
requires two loop RG running from $M_G$ to $M_Z$ and one loop
threshold corrections at the weak scale; GUT scale threshold
corrections are now needed to precisely fit the low energy data.} }
\end{center}
\end{figure}

At present gauge coupling unification within SUSY GUTs works
extremely well.  Exact unification at $M_G$, with two loop
renormalization group running from $M_G$ to $M_Z$,  and one loop
threshold corrections at the weak scale, fits to within 3 $\sigma$
of the present precise low energy data.  A small threshold
correction at $M_G$ ($\epsilon_3 \sim $ - 3\%  to - 4\%) is
sufficient to fit the low energy data
precisely~\cite{Lucas:1995ic,dmr,Alciati:2005ur}.\footnote{This
result implicitly assumes universal GUT boundary conditions for soft
SUSY breaking parameters at $M_G$. In the simplest case we have a
universal gaugino mass $M_{1/2}$, a universal mass for squarks and
sleptons $m_{16}$ and a universal Higgs mass $m_{10}$, as motivated
by $SO(10)$.  In some cases, threshold corrections to gauge coupling
unification can be exchanged for threshold corrections to soft SUSY
parameters.  For a recent review, see \cite{pdg}.} See Fig.
\ref{fig:guts.2loop}. This may be compared to non-SUSY GUTs where
the fit misses by $\sim$ 12 $\sigma$ and a precise fit requires new
weak scale states in incomplete GUT multiplets or multiple GUT
breaking scales.\footnote{Non-SUSY GUTs with a more complicated
breaking pattern can still fit the data. For a recent review, see
\cite{pdg}.} When GUT threshold corrections are considered, then all
three gauge couplings no longer meet at a point.  Thus we have some
freedom in how we define the GUT scale.   We choose to define the
GUT scale as the point where $\alpha_1(M_G) = \alpha_2(M_G) \equiv
\tilde \alpha_G$ and $\alpha_3(M_G) = \tilde \alpha_G \; (1 +
\epsilon_3)$. The threshold correction $\epsilon_3$ is a logarithmic
function of all states with mass of order $M_G$ and $\tilde \alpha_G
= \alpha_G + \Delta$ where $\alpha_G$ is the GUT coupling constant
above $M_G$ and $\Delta$ is a one loop threshold correction. To the
extent that gauge coupling unification is perturbative, the GUT
threshold corrections are small and calculable. This presumes that
the GUT scale is sufficiently below the Planck scale or any other
strong coupling extension of the GUT, such as a strongly coupled
string theory.

In four dimensional SUSY GUTs, the threshold correction $\epsilon_3$
receives a positive contribution from Higgs doublets and triplets.
Note, the Higgs contribution is given by $\epsilon_3 = {3
\tilde\alpha_G \over 5\pi}\log |{ \tilde{M_t} \ \gamma \over M_G}|$
where $\tilde{M_t}$ is the effective color triplet Higgs mass
(setting the scale for dimension 5 baryon and lepton number
violating operators) and $\gamma = \lambda_b/\lambda_t$ at $M_G$.
Obtaining $\epsilon_3 \sim - 3\%$ (with $\gamma = 1$) requires
$\tilde{M_t} \sim 10^{14}$ GeV. Unfortunately this value of
$\tilde{M_t}$ is now excluded by the non-observation of proton
decay.\footnote{With $\gamma \sim m_b/m_t$, we need $\tilde{M_t}
\sim M_G$ which is still excluded by proton decay.}  In fact, as we
shall now discuss, the dimension 5 operator contribution to proton
decay requires $\tilde{M_t}$ to be greater than $M_G$.  In this case
the Higgs contribution to $\epsilon_3$ is positive. Thus a larger,
negative contribution must come from the GUT breaking sector of the
theory. This is certainly possible in specific SO(10)
~\cite{so10breaking} or SU(5) ~\cite{af} models, but it is clearly a
significant constraint on the 4d GUT-breaking sector of the theory.

Baryon number is necessarily violated in any GUT~\cite{grs}.  In any
grand unified theory, nucleons decay via the exchange of gauge
bosons with GUT scale masses, resulting in dimension 6 baryon number
violating operators suppressed by $(1/M_G^2)$.  The nucleon lifetime
is calculable and given by $\tau_N \propto M_G^4/(\alpha_G^2 \;
m_p^5)$.  The dominant decay mode of the proton (and the baryon
violating decay mode of the neutron), via gauge exchange, is $p
\rightarrow e^+ \; \pi^0$ ($n \rightarrow e^+ \; \pi^-$).  In any
simple gauge symmetry, with one universal GUT coupling and scale
($\alpha_G, \; M_G$), the nucleon lifetime from gauge exchange is
calculable.  Hence, the GUT scale may be directly observed via the
extremely rare decay of the nucleon.  The present experimental
bounds come from Super-Kamiokande. We discuss these results shortly.
In SUSY GUTs, the GUT scale is of order $3\times 10^{16}$ GeV, as
compared to the GUT scale in non-SUSY GUTs which is of order
$10^{15}$ GeV. Hence the dimension 6 baryon violating operators are
significantly suppressed in SUSY GUTs~\cite{drw} with $\tau_p \sim
10^{34 - 38}$ yrs.

\begin{figure}[t!]
\hspace{1.0in}
\begin{center}
\begin{minipage}{3in}
\epsfig{file=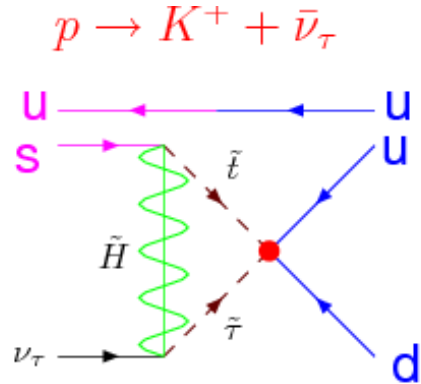,height=2.0in}
\end{minipage}
\caption{\label{fig:proton.susy} {\small The contribution of
dimension 5 operators to proton decay.  The red dot denotes the
dimension 5 vertex. } }
\end{center}
\end{figure}

However, in SUSY GUTs there are additional sources for baryon number
violation -- dimension 4 and 5 operators~\cite{bviol}.  The
dimension 4 operators violate baryon number or lepton number,
respectively, but not both.  The nucleon lifetime is extremely short
if both types of dimension 4 operators are present in the low energy
theory.  However both types can be eliminated by requiring R parity
\cite{rparity}. R parity distinguishes Higgs multiplets from
ordinary families. In $SU(5)$, Higgs and quark/lepton multiplets
have identical quantum numbers; while in $E(6)$, Higgs and families
are unified within the fundamental ${\bf 27}$ representation. Only
in SO(10) are Higgs and ordinary families distinguished by their
gauge quantum numbers. In what follows we shall assume that R parity
is a good symmetry and neglect further discussion of Dimension 4
baryon and/or lepton number violating operators.   As a consequence,
the lightest SUSY particle [LSP] is stable and is an excellent dark
matter candidate.

The dimension 5 operators have a dimensionful coupling of order
($1/M_G$).  Dimension 5 baryon number violating operators may be
forbidden at tree level by symmetries in $SU(5)$, etc. These
symmetries are typically broken however by the VEVs responsible for
the color triplet Higgs masses. Consequently these dimension 5
operators are generically generated via color triplet Higgsino
exchange.  Hence, the color triplet partners of Higgs doublets must
necessarily obtain mass of order the GUT scale.  The dominant decay
modes from dimension 5 operators are $p \rightarrow K^+ \; \bar \nu
\;\; (n \rightarrow K^0 \; \bar \nu)$. See Fig.
\ref{fig:proton.susy}. This is due to a simple symmetry argument;
the operators $(Q_i \; Q_j\; Q_k\; L_l), \;\; (u^c_i \; u^c_j\;
d^c_k\; e^c_l)$ (where $i,\; j,\; k,\; l = 1,2,3$ are family indices
and color and weak indices are implicit) must be invariant under
$SU(3)_C$ and $SU(2)_L$.  As a result their color and weak doublet
indices must be anti-symmetrized.  However since these operators are
given by bosonic superfields, they must be totally symmetric under
interchange of all indices. Thus the first operator vanishes for $i
= j = k$ and the second vanishes for $i = j$.  Hence a second or
third generation member must exist in the final state~\cite{drw2}.

Super-Kamiokande bounds on the proton lifetime severely constrain
these dimension 6 and 5 operators with $\tau_{(p \rightarrow e^+
\pi^0)} > 5.0 \times 10^{33}$ yrs (79.3 ktyr exposure), $\tau_{(n
\rightarrow e^+ \pi^-)} > 5 \times 10^{33}$ yrs (61 ktyr), and
$\tau_{(p \rightarrow  K^+ \bar \nu)} > 2.3 \times 10^{33}$ yrs (92
ktyr), $\tau_{(n \rightarrow K^0 \bar \nu)} > 1.3 \times 10^{32}$
yrs (92 ktyr) at (90\% CL) based on the listed
exposures~\cite{superk}. These constraints are now sufficient to
rule out minimal SUSY $SU(5)$~\cite{murayama}.   Non-minimal Higgs
sectors in $SU(5)$ or $SO(10)$ theories still survive~\cite{dmr,af}.
The upper bound on the proton lifetime from these theories are
approximately a factor of 5 above the experimental bounds. They are,
however, being pushed to their theoretical limits.  Hence if SUSY
GUTs are correct, nucleon decay must be seen soon.

\subsection{4D  vs.  Orbifold GUTs}

\begin{figure}[t!]
\hspace{-.70in}
\begin{center}
\begin{minipage}{4in}
\epsfig{file=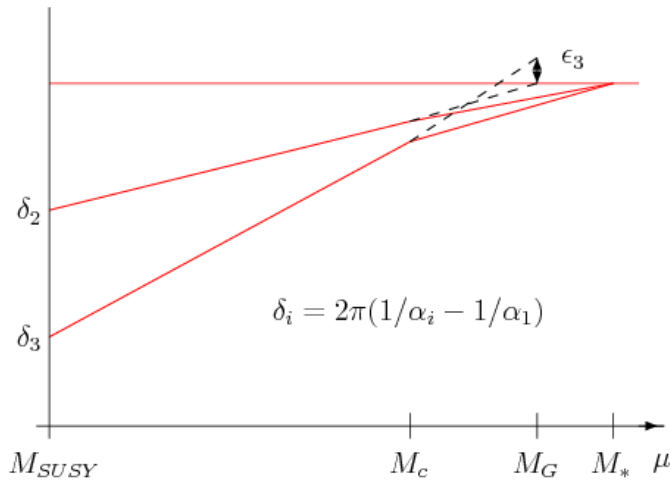,height=2.5in}
\end{minipage}
\caption{\label{fig:oguts} {\small One loop RG running in orbifold
GUTs from Kim and Raby \cite{kawamura}.} }
\end{center}
\end{figure}

Orbifold compactification of the heterotic
string~\cite{heterotic,Witten:2001bf,krz,string.oguts}, and recent
field theoretic constructions known as orbifold
GUTs~\cite{kawamura}, contain grand unified symmetries realized in 5
and 6 dimensions. However, upon compactifying all but four of these
extra dimensions, only the MSSM is recovered as a symmetry of the
effective four dimensional field theory. These theories can retain
many of the nice features of four dimensional SUSY GUTs, such as
charge quantization, gauge coupling unification and sometimes even
Yukawa unification; while at the same time resolving some of the
difficulties of 4d GUTs, in particular problems with unwieldy Higgs
sectors necessary for spontaneously breaking the GUT symmetry, and
problems with doublet-triplet Higgs splitting or rapid proton decay.

In five or six dimensional orbifold GUTs, on the other hand, the
``GUT scale" threshold correction comes from the Kaluza-Klein modes
between the compactification scale, $M_c$, and the effective cutoff
scale $M_*$.\footnote{In string theory, the cutoff scale is the
string scale.}  Thus, in orbifold GUTs,  gauge coupling unification
at two loops is only consistent with the low energy data with a
fixed value for $M_c$ and $M_*$.\footnote{It is interesting to note
that a ratio $M_*/M_c \sim 100$, needed for gauge coupling
unification to work in orbifold GUTs is typically the maximum value
for this ratio consistent with perturbativity~\cite{dienes}.  In
addition, in orbifold GUTs brane-localized gauge kinetic terms may
destroy the successes of gauge coupling unification.  However,  for
values of $M_*/M_c = M_* \pi R \gg 1$ the unified bulk gauge kinetic
terms can dominate over the brane-localized
terms~\cite{hallnomura}.} Typically, one finds $M_c < M_G (= 3
\times 10^{16}$ GeV) $< M_*$, where $M_G$ is the 4d GUT scale. Since
the grand unified gauge bosons, responsible for nucleon decay, get
mass at the compactification scale,  the result $M_c < M_G$ for
orbifold GUTs has significant consequences for nucleon decay.   See
Fig. \ref{fig:oguts}.

Orbifold GUTs and string theories contain grand unified symmetries
realized in higher dimensions.   In the process of compactification
and GUT symmetry breaking, color triplet Higgs states are removed
(projected out of the massless sector of the theory). In addition,
the same projections typically rearrange the quark and lepton states
so that the massless states which survive emanate from different GUT
multiplets.  In these models, proton decay due to dimension 5
operators can be severely suppressed or eliminated completely.
However, proton decay due to dimension 6 operators may be enhanced,
since the gauge bosons mediating proton decay obtain mass at the
compactification scale, $M_c$, which is less than the 4d GUT scale,
or suppressed, if the states of one family come from different
irreducible representations. Which effect dominates is a model
dependent issue. In some complete 5d orbifold GUT
models~\cite{complete,Alciati:2005ur} the lifetime for the decay
$\tau(p \rightarrow e^+ \pi^0)$ can be near the excluded bound of $5
\times 10^{33}$ years with, however, large model dependent and/or
theoretical uncertainties.  In other cases, the modes $p \rightarrow
K^+ \bar \nu$ and $p \rightarrow K^0 \mu^+$ may be
dominant~\cite{Alciati:2005ur}.  To summarize, in either 4d or
orbifold string/field theories, nucleon decay remains a premier
signature for SUSY GUTs. Moreover, the observation of nucleon decay
may distinguish extra-dimensional orbifold GUTs from four
dimensional ones.

\subsection{Heterotic string/orbifold GUTs}

In recent years there has been a great deal of progress in
constructing three and four family models in Type IIA string theory
with intersecting D6 branes~\cite{cvetic}. Although these models can
incorporate SU(5) or a Pati-Salam symmetry group in four dimensions,
they typically have problems with gauge coupling unification. In the
former case this is due to charged exotics which affect the RG
running, while in the latter case the SU(4)$\times$SU(2)$_L
 \times$SU(2)$_R$ symmetry never unifies. Note, heterotic string
theory models also exist whose low energy effective 4d field theory
is a SUSY GUT~\cite{stringgut}. These models have all the virtues
and problems of 4d GUTs. Finally, many heterotic string models have
been constructed with the standard model gauge symmetry in 4d and no
intermediate GUT symmetry in less than 10d. Recently some minimal 3
family supersymmetric models have been
constructed~\cite{Cleaver:1999cj,Braun:2005ux}. These theories may
retain some of the symmetry relations of GUTs, however the
unification scale would typically be the string scale, of order $5
\times 10^{17}$ GeV, which is inconsistent with low energy data.   A
way out of this problem was discovered in the context of the
strongly coupled heterotic string, defined in an effective 11
dimensions~\cite{Witten:1996mz}.   In this case the 4d Planck scale
(which controls the value of the string scale) now unifies with the
GUT scale.
\begin{figure}[t!]
\hspace{-.70in}
\begin{center}
\begin{minipage}{4in}
\epsfig{file=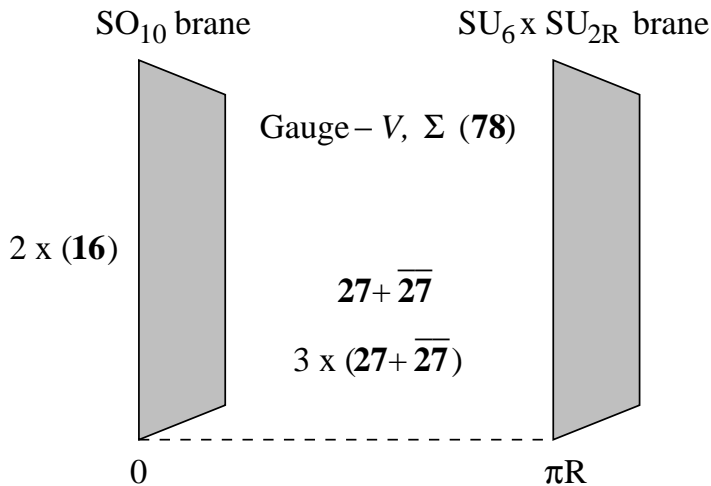,height=2.5in}
\end{minipage}
\caption{\label{fig:e6ogut} {\small Schematic of an E(6) orbifold
GUT in 5 dimensions with details in the text.} }
\end{center}
\end{figure}

Recently a new paradigm has been proposed for using orbifold GUT
intuition for constructing heterotic string models \cite{krz}.
Consider an $E(6)$ orbifold GUT defined in 5 dimensions, where the
5th dimension is a line segment from $y = 0$ to $\pi R$, as given in
Fig. \ref{fig:e6ogut}.    In this theory the bulk contains the gauge
hypermultiplet $V,  \ \Sigma$ (in 4 dimensional N=1 SUSY notation)
and four {\bf 27} dimensional hypermultiplets.   The orbifold
parities $Z_2 \times Z_2^\prime$ leave an $SO(10)$ invariant brane
at $y = 0$ and an $SU(6) \times SU(2)_R$ invariant brane at $y = \pi
R$.   The overlap between these two localized symmetries is the low
energy symmetry of the theory, which in this case is Pati-Salam.
Note the theory also contains massless Higgs bosons necessary to
spontaneously break PS to the standard model. In orbifold GUT
language the compactification scale is given by $M_c = 1/\pi R$ and
the theory is cut-off at a scale $M_* \gg M_c$. In this theory we
have the two light families of quarks and leptons residing on the
$SO(10)$ brane, while the Higgs doublets and third family are
located in the bulk.  Finally the compactification scale is of order
$7 \times 10^{15}$ GeV.   Thus proton decay via dimension 6
operators is enhanced,   while proton decay via dimension 5
operators is forbidden.

\begin{figure}[t!]
\hspace{-.70in}
\begin{center}
\begin{minipage}{5in}
\epsfig{file=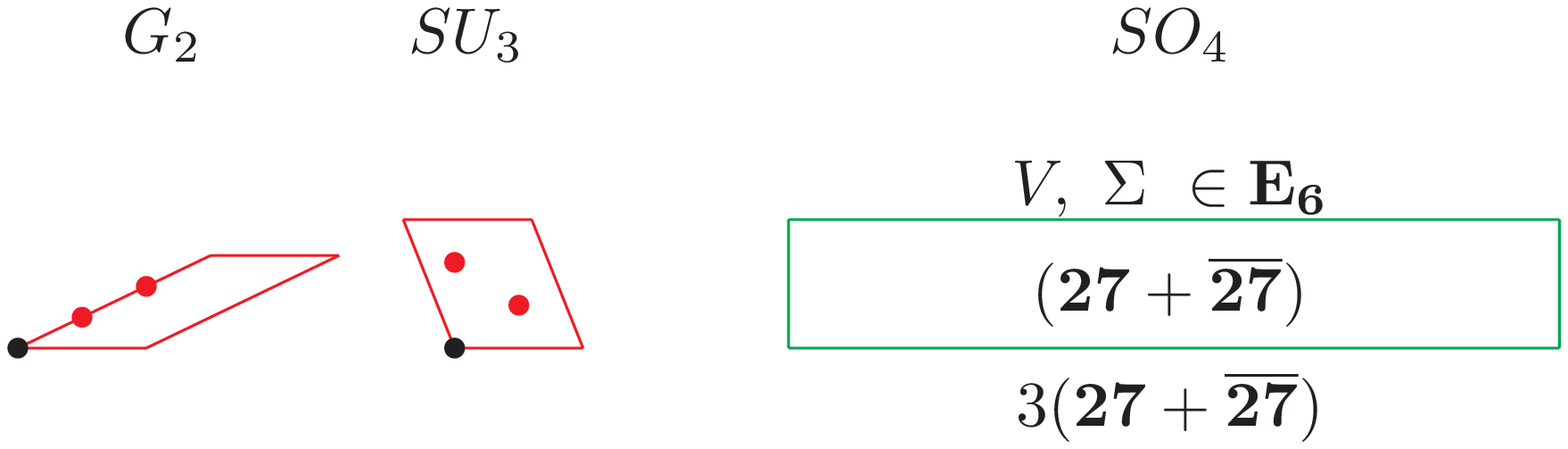,height=1.5in}
\end{minipage}
\caption{\label{fig:e6heterotic} {\small The bulk states in an
effective 5D orbifold GUT obtained from the heterotic string.  More
details in the text.} }
\end{center}
\end{figure}
\begin{figure}[t!]
\hspace{-.70in}
\begin{center}
\begin{minipage}{5in}
\epsfig{file=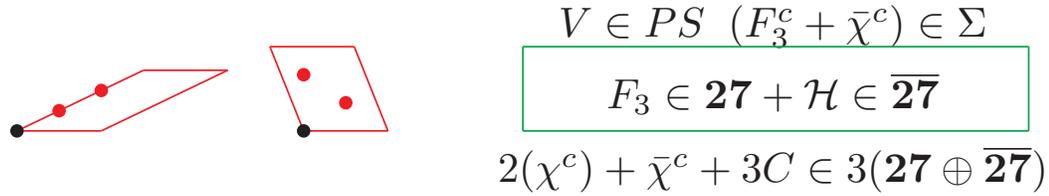,height=1.0in}
\end{minipage}
\caption{\label{fig:PS} {\small The bulk states in effective 5D
orbifold GUT after the $Z_2$ twist and Wilson line.   More details
in the text.} }
\end{center}
\end{figure}

Now let's discuss how one obtains this model from the $E(8) \times
E(8)$ heterotic string in 10 dimensions \cite{krz}.   One
compactifies 6 dimensions on 3 two dimensional torii defined by the
root lattice of $G_2 \times SU(3) \times SO(4)$ given in Fig.
\ref{fig:e6heterotic}. One must also mod out by a $Z_2 \times Z_3$
orbifold symmetry of the lattice.   Consider first $(T^2)^3/Z_3$
where the $Z_3$ twist is also embedded in the $E(8) \times E(8)$
group lattice via a shift vector $V_3$ and a Wilson line $W_3$ in
the $SU(3)$ torus.   Note, the $Z_3$ twist only acts on the $G_2
\times SU(3)$ torii,  leaving the $SO(4)$ torus untouched.   As a
result of the orbifolding, the $E(8) \times E(8)$ symmetry is broken
to $E(6) \times$ an $SO(10) \times U(1)^5$ hidden sector gauge
symmetry.   We will not consider the hidden sector further in the
discussion.   The untwisted sector of the string contains the
following massless states,  $V, \ \Sigma$ in the adjoint of $E_6$
and one {\bf 27} dimensional hypermultiplet.   Three more {\bf 27}
dimensional hypermultiplets sit at the trivial fixed point in the
$SU(3)$ torus and one each at the 3 $G_2$ fixed points.   However at
this level the massless string fields can be described by an
effective 6 dimensional theory.   Out of the 6 compactified
dimensions we take 5 to be of order the string scale and one much
larger than the string scale, as in Fig. \ref{fig:e6heterotic}.   In
this case, we now have the first step in the 5 dimensional orbifold
GUT described earlier.    In order to break the $E(6)$ gauge
symmetry we now apply the $Z_2$ twist, acting on the $G_2 \times
SO(4)$ torii, and embedded into the $E(8) \times E(8)$ gauge lattice
by a shift vector $V_2$ and a Wilson line, $W_2$, along the long
axis of the $SO(4)$ torus.   As a result we find the massless sector
shown in Fig. \ref{fig:PS}.

\begin{figure}[t!]
\hspace{-.70in}
\begin{center}
\begin{minipage}{5in}
\epsfig{file=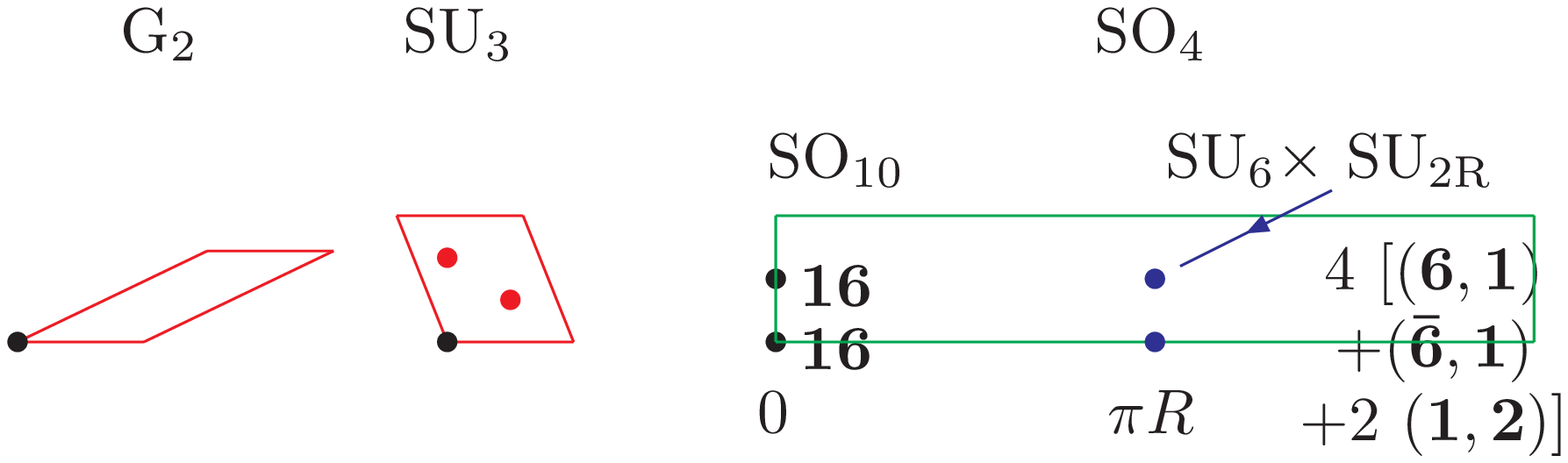,height=1.5in}
\end{minipage}
\caption{\label{fig:T1sector} {\small The twisted sector states in
the $T_1$ twisted sector.} }
\end{center}
\end{figure}

The shift $V_2$ can be identified with the first orbifold parity
leaving two $SO(10)$ invariant fixed points in the $SO(4)$ torus.
While the combination $V_2 + W_2$ leaves two $SU(6) \times SU(2)_R$
invariant fixed point at the opposite side of the orbifold. See Fig.
\ref{fig:T1sector} which describes the $T_1$ twisted sector of the
string.   Note we also find two complete {\bf 16} dimensional
representations of $SO(10)$ residing, one on each of the $SO(10)$
fixed points.

\begin{figure}[t!]
\hspace{-.30in}
\begin{center}
\begin{minipage}{5in}
\epsfig{file=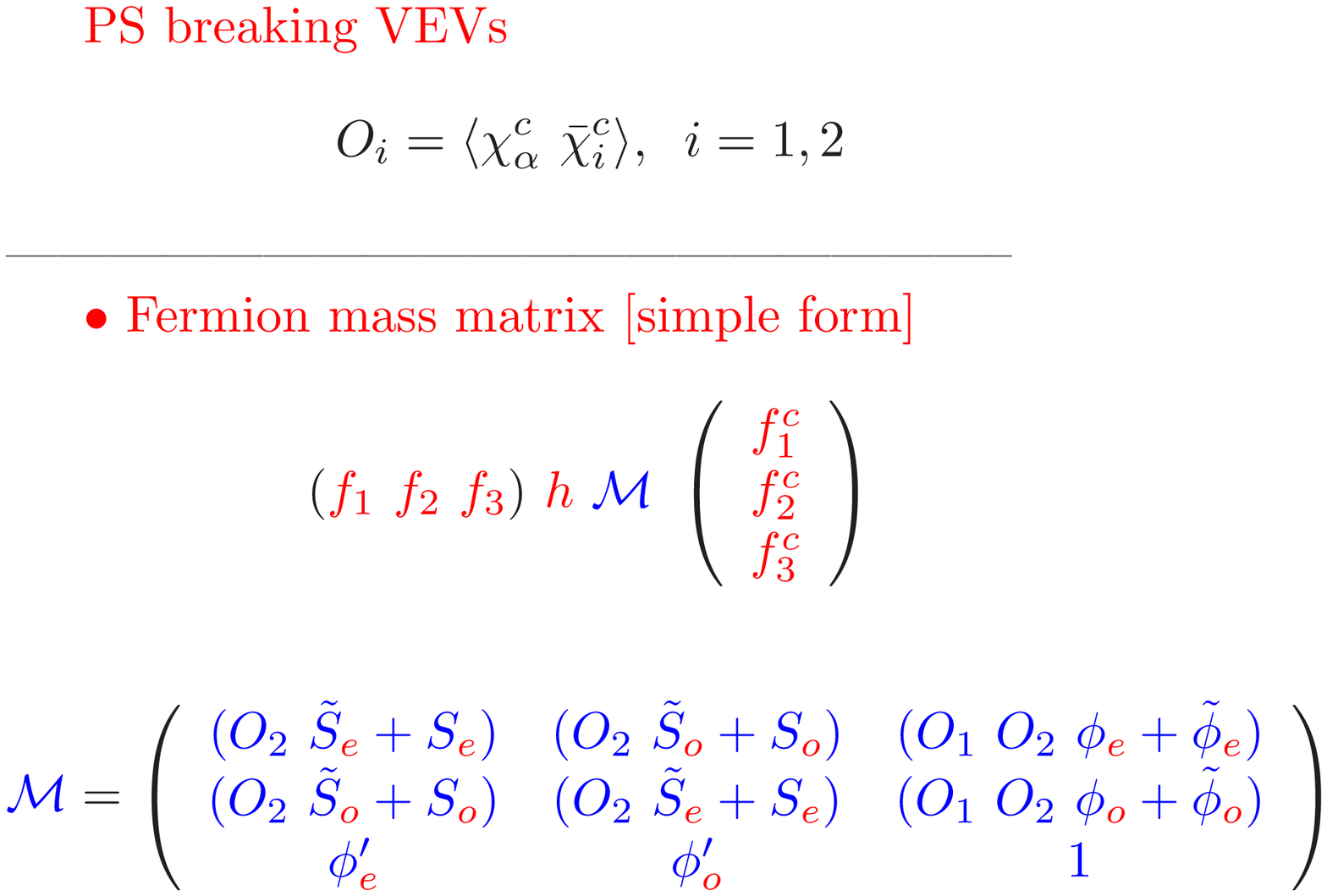,height=3.0in}
\end{minipage}
\caption{\label{fig:d4yukawa} {\small } }
\end{center}
\end{figure}

\begin{figure}[t!]
\hspace{-.70in}
\begin{center}
\begin{minipage}{5in}
\epsfig{file=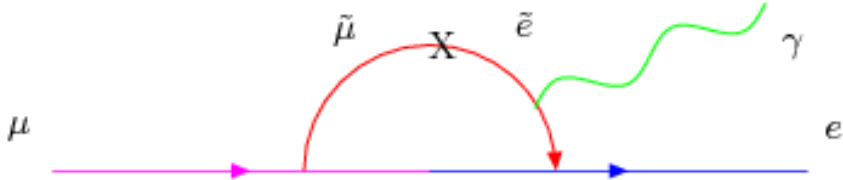,height=1.0in}
\end{minipage}
\caption{\label{fig:muegamma} {\small One loop contribution to the
process $\mu \rightarrow e \ \gamma$ in a SUSY flavor basis defined
by lepton mass eigenstates, BUT not necessarily slepton mass
eigenstates.} }
\end{center}
\end{figure}

As a consequence of the string selection rules we find a $D_4$
family symmetry,  where $D_4 = \{ \pm 1,  \pm \sigma_1, \pm
\sigma_3, \pm i \sigma_2 \}$.   The two light families transform as
the fundamental doublet,  where the action of $\sigma_1 \
[\sigma_3]$ is given by $(f_1, \ f_1^c) \leftrightarrow (f_2, \
f_2^c) \; [ (f_2, \ f_2^c) \leftrightarrow - (f_2, \ f_2^c)]$.   The
third family, given by the PS multiplets $f_3 =  (4, 2, 1)$ and
$f_3^c = (\bar 4, 1, \bar 2)$, and the Higgs doublet, ${\cal H} =
(1, \bar 2, 2)$, transform as $D_4$ singlets.    As a result of the
$D_4$ family symmetry, we find that the effective Yukawas for the
theory are constrained to be of the form given in Fig.
\ref{fig:d4yukawa},  where the terms $\tilde S_{e,o}$ are products
of vevs of standard model singlets, even or odd under $\pm \sigma_3$
and $O_{1,2}$ are the vevs of PS breaking Higgs multiplets.
Note the non-abelian family symmetry $D_4$ is a discrete subgroup of
$SU(2)$.   Unbroken $D_4$ symmetry requires degenerate masses for
squarks and sleptons of the light two families, hence suppressing
flavor violating processes such as $\mu \rightarrow e \gamma$ (Fig.
\ref{fig:muegamma}).

\begin{figure}[t!]
\hspace{-.70in}
\begin{center}
\begin{minipage}{5in}
\epsfig{file=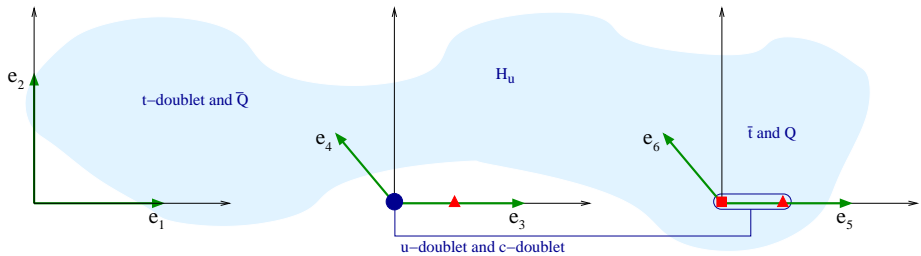,height=1.3in}
\end{minipage}
\caption{\label{fig:z2xz4} {\small Some of the states obtained in a
$(T^2)^3/(Z_2 \times Z_4)$ orbifold compactification of the E(8)
$\times$ E(8) heterotic string to the Standard Model gauge symmetry.
 See test for more details.} }
\end{center}
\end{figure}
We are now performing a search in the ``string landscape" for
standard model theories in four dimensions compactified on a
$(T^2)^3/(Z_2 \times Z_3)$ or $(T^2)^3/(Z_2 \times Z_4)$ orbifold
\cite{rvw}.  These can be obtained with a combination of shift
vectors  V and up to 3 Wilson lines.  Yes, we are searching for the
MSSM in 4D without including GUTs.

Consider, for example,  the case $(T^2)^3/(Z_2 \times Z_4)$.  The
search strategy is to look for models with gauge group $SU(3) \times
SU(2) \times U(1)^n \times$ a possible non-abelian hidden sector
group. At the first step, we identify a three family model as one
which has 3 more $({\bf 3,2})$ than $({\bf \bar 3, 2})$, 6 more
$({\bf \bar 3, 1})$ than their conjugates and at least 5 $({\bf 1,
2})$.  The last condition guarantees that we have 3 families of
lepton doublets and at least two higgs doublets.  There are 120
inequivalent possibilities for the shift vectors $V_2, \ V_4$.  Then
with either one or two Wilson lines we find 500,000 possible models.

However at this level, we have yet to identify the weak hypercharge
as one linear combination of all the $U(1)$ charges. We find that,
in general, the models satisfying this initial cut have non-standard
model hypercharge assignments and in most cases, chiral exotics,
i.e. fractionally charged states which cannot obtain mass. We then
needed a search strategy to find standard model hypercharge
assignments for quarks and leptons.   Aha,  if we require that the 3
families come in complete $SU(5)$ or PS multiplets, at an
intermediate state AND identified hypercharge as $Y \in SU(5)$ or $Y
=  (B-L) + 2 T_{3R} \in PS$,  then we found 3 family models with
only vector-like exotics.   Thus the hypothesis we are now testing
is whether quarks and leptons coming from complete GUT multiplets
are necessary for {\bf charge quantization}.  Our preliminary
results are as follows, we find \begin{enumerate} \item 6 models via
PS $\rightarrow$ SM, and \item 40  models via $SU(5) \rightarrow$ SM
.\end{enumerate} If we now look for models with 3 families of quarks
and leptons with Pati – Salam symmetry in 4 dimensions and the
necessary Higgs bosons to break PS to the standard model AND no
chiral exotics, we find about 220 models.

One example of a three family model with vector-like exotics in $Z_2
\times Z_4$ orbifold is given in Fig. \ref{fig:z2xz4}.  The top
quark doublet, left-handed anti-top and $H_u$ reside in the bulk and
has a tree level Yukawa coupling.   The left-handed anti-bottom and
tau lepton resides on twisted sector fixed points and their Yukawa
couplings come only at order $S^2$.   Finally the light two families
come on two distinct fixed points in the twisted sector,  although
the description is a bit more complicated than our earlier model. We
have not analyzed these models in any great detail, so for example,
we do not know the complete family symmetry of the theory or whether
all the vector-like exotics can get a large mass.

\section{Minimal SO$_{10}$ SUSY Model [MSO$_{10}$SM] and Large $\tan\beta$}

Let me first define what I mean by the [MSO$_{10}$SM]
\cite{bdr,Dermisek:2003vn,Dermisek:2006dc}. Quarks and leptons of
one family reside in the $\bf 16$ dimensional representation, while
the two Higgs doublets of the MSSM reside in a single $\bf 10$
dimensional representation.  For the third generation we assume the
minimal Yukawa coupling term given by $ {\bf \lambda \ 16 \ 10 \ 16
}. $ On the other hand, for the first two generations and for their
mixing with the third, we assume a hierarchical mass matrix
structure due to effective higher dimensional operators. Hence the
third generation Yukawa couplings satisfy $\lambda_t = \lambda_b =
\lambda_\tau = \lambda_{\nu_\tau} = {\bf \lambda}$.

Soft SUSY breaking parameters are also consistent with $SO_{10}$
with (1) a universal gaugino mass $M_{1/2}$, (2) a universal squark
and slepton mass $m_{16}$,\footnote{$SO_{10}$ does not require all
sfermions to have the same mass. This however may be enforced by
non--abelian family symmetries or possibly by the SUSY breaking
mechanism.} (3) a universal scalar Higgs mass $m_{10}$, and (4) a
universal A parameter $A_0$. In addition we have the supersymmetric
(soft SUSY breaking) Higgs mass parameters $\mu$ ($B \mu$).  $B \mu$
may, as in the CMSSM, be exchanged for $\tan\beta$. Note, not all of
these parameters are independent. Indeed, in order to fit the low
energy electroweak data, including the third generation fermion
masses, it has been shown that $A_0, \ m_{10}, \ m_{16}$ must
satisfy the constraints~\cite{bdr}
\begin{eqnarray} A_0 \approx - 2 \ m_{16}; & m_{10} \approx \sqrt{2} \ m_{16}  
\label{eq:constraint1}
\\
m_{16} > 1.2 \; {\rm TeV}; & \mu, \ M_{1/2} \ll m_{16} 
\label{eq:constraint2} \end{eqnarray} with \begin{equation}
\tan\beta \approx 50. \label{eq:tanbeta} \end{equation}

\subsection{Dark matter and WMAP data}

\begin{figure}[t!]
\begin{center}
\begin{minipage}{3.5in}
\epsfig{file=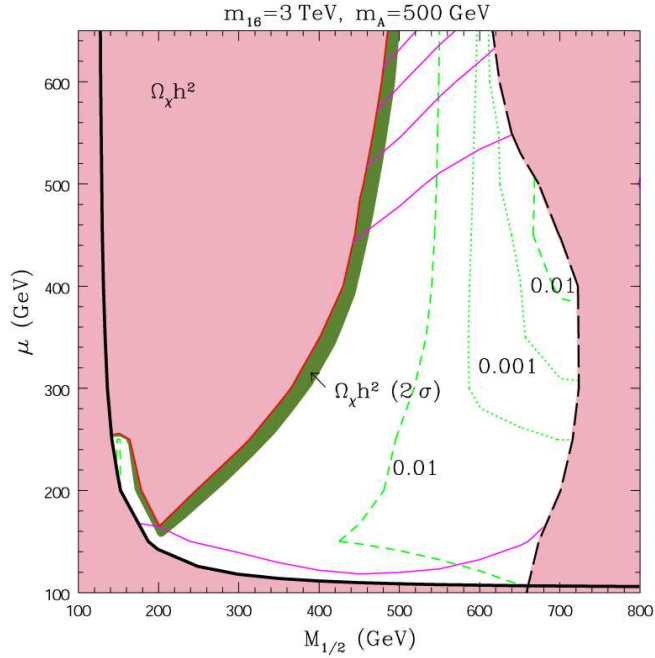,height=3.5in}
\end{minipage}
\caption{\label{fig:wmap} {\small Neutralino dark matter abundance
from Ref. \cite{Dermisek:2003vn}.} }
\end{center}
\end{figure}

\begin{figure}[t!]
\hspace{-.70in}
\begin{center}
\begin{minipage}{4in}
\epsfig{file=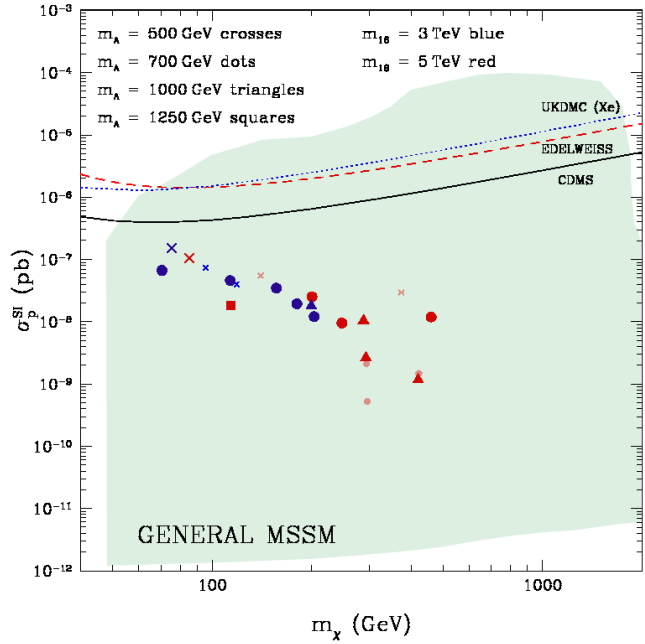,height=3.5in}
\end{minipage}
\caption{\label{fig:sip} {\small Spin-independent neutralino -
proton scattering cross-section from Ref. \cite{Dermisek:2003vn}.} }
\end{center}
\end{figure}

The model is assumed to have a conserved R parity, thus the LSP is
stable and an excellent dark matter candidate.   In our case the LSP
is a neutralino; roughly half higgsino and half gaugino.   The
dominant annihilation channel is via an s-channel CP odd Higgs, {\bf
A}. In the large $\tan\beta$ limit,  {\bf A} is a wide resonance
since its coupling to bottom quarks and taus is proportional to
$\tan\beta$.   In Fig. \ref{fig:wmap} we give the fit to WMAP dark
matter abundance as a function of $\mu, \ M_{1/2}$ for $m_{16} = 3$
TeV and $m_A = 700$ GeV \cite{Dermisek:2003vn}.  The green (darker
shaded) area is consistent with WMAP data.  While to the left the
dark matter abundance is too large and to the right (closer to the
{\bf A} peak) it is too small.  In Fig. \ref{fig:sip} we give the
spin independent neutralino-proton cross-section relevant for dark
matter searches. All the points give acceptable fits for top, bottom
and tau masses, as well as $\alpha_i, \ i = 1,2,3$ at $M_Z$ with
$Br(B_s \rightarrow \mu^+ \mu^-) \le 5 \times 10^{-7}$.  The large
dots also satisfy $Br(B_s \rightarrow \mu^+ \mu^-) \le 2 \times
10^{-7}$.

\begin{figure}[t!]
\hspace{-0.70in}
\begin{center}
\begin{minipage}{6in}
\epsfig{file=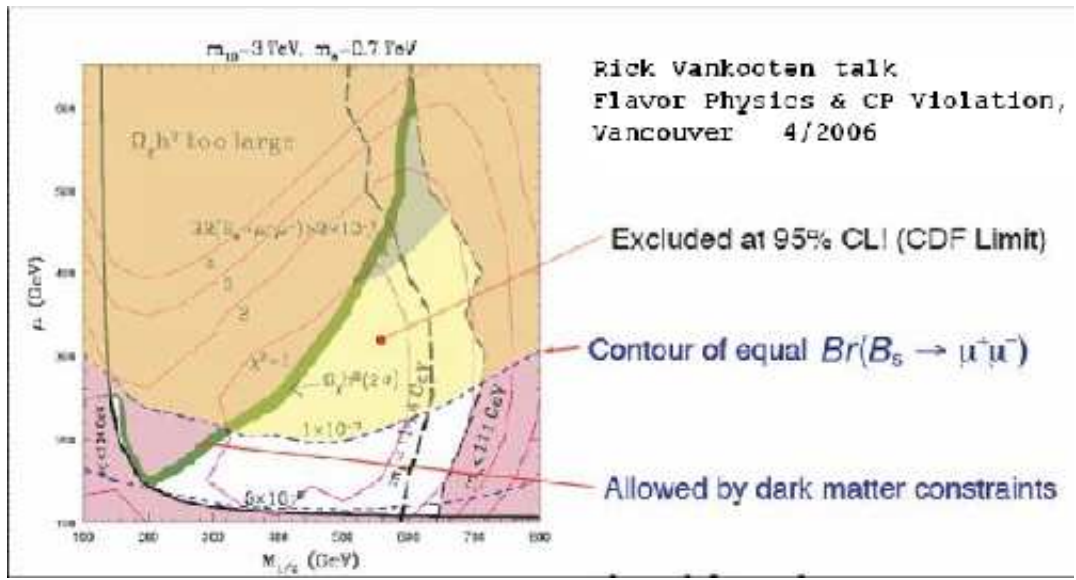,height=3.0in}
\end{minipage}
\caption{\label{fig:mh-bsmm} {\small Graph from Ref.
\cite{Dermisek:2003vn} with annotations by Rick Vankooten. } }
\end{center}
\end{figure}

\subsection{B physics and both the CP odd and light Higgs masses}

It is well known that the large $\tan\beta$ region of the MSSM has
many interesting consequences for B physics.    Let us just consider
some interesting examples.   We shall consider the branching ratio
$Br(B_s \rightarrow \mu^+ \mu^-)$,  the mass splitting $\Delta
M_{B_s}$, the branching ratio \linebreak $Br(B_u \rightarrow \tau
\bar \nu)$ and the related processes $B \rightarrow X_s \gamma$ and
$B \rightarrow X_s \ l^+ \ l^-$.

\subsubsection{$Br(B_s \rightarrow \mu^+ \mu^-)$}

In Fig. \ref{fig:mh-bsmm} we give the contours for the branching
ratio $Br(B_s \rightarrow \mu^+ \mu^-)$ plotted as a function of
$\mu$ and $M_{1/2}$ for fixed $m_{16} = 3$ TeV and fixed CP odd
Higgs mass, $m_A = 700$ GeV.  The most recent 'preliminary' bounds
from CDF \cite{cdf-bsmm} and DZero \cite{d0-bsmm} are $Br( B_s
\rightarrow \mu^+ \mu^- ) < 1.0 \times 10^{-7}$ @ 95\% CL for CDF
with 780 pb$^{-1}$ data and $< 2.3 \times 10^{-7}$ @ 95\% CL for
DZero with 700 pb$^{-1}$ data. Thus only the region below the dashed
blue contour $1.0 \times 10^{-7}$ is allowed.  Note the black dashed
vertical contour on the right. This is the light Higgs mass bound.
The Higgs mass decreases as $M_{1/2}$ increases.

\begin{figure}[t!]
\hspace{-.70in}
\begin{center}
\begin{minipage}{4in}
\epsfig{file=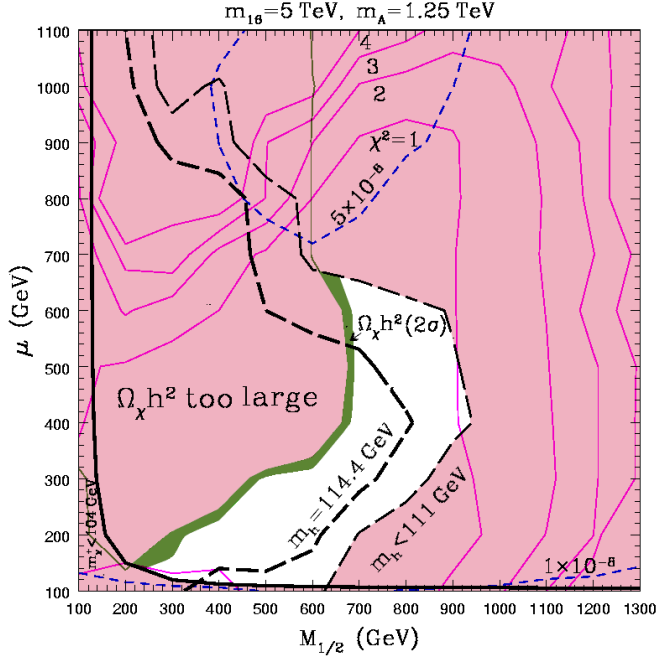,height=3.5in}
\end{minipage}
\caption{\label{fig:drrr2-m05000-mA1250} {\small Neutralino
abundance and LEP Higgs bound for the CP odd Higgs mass, $m_A =
1.25$ TeV.  Contours of constant $Br(B_s \rightarrow \mu^+ \mu^-)$
are dashed blue lines.} }
\end{center}
\end{figure}

\begin{figure}[t!]
\hspace{-.70in}
\begin{center}
\begin{minipage}{4in}
\epsfig{file=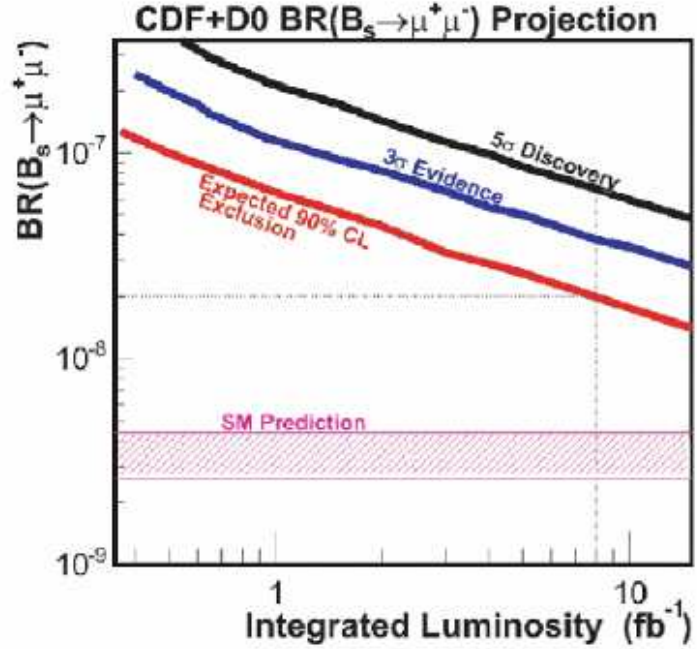,height=3.5in}
\end{minipage}
\caption{\label{fig:bsmm} {\small Sensitivity of Fermilab
experiments to $Br(B_s \rightarrow \mu^+ \mu^-)$.  Copied from talk
by Rick Vankooten at Flavor Physics \& CP Violation,  Vancouver,
April 2006. } }
\end{center}
\end{figure}

The amplitude for the process $B_s \rightarrow \mu^+ \mu^-$ is
dominated by the s-channel CP odd Higgs exchange and the branching
ratio scales as $\tan\beta^6/m_A^4$.   Thus as $m_A$ increases the
branching ratio can be made arbitrarily small.   On the other hand,
we find that as we increase $m_A$, we must necessarily increase
$M_{1/2}$ in order to satisfy  WMAP data. Since in order to have
sufficient annihilation we must approximately satisfy  $m_A \sim 2
m_\chi$ and $m_\chi \propto M_{1/2}$. However, the light Higgs mass
decreases as $M_{1/2}$ increases and hence there is an upper bound,
$m_A(MAX)$, such that WMAP and the light Higgs mass lower bound are
both satisfied.   We find \cite{Dermisek:2003vn} $m_A(MAX) \sim 1.3$
TeV and as a result $Br(B_s \rightarrow \mu^+ \mu^-)(MIN) \sim 1
\times 10^{-8}$ (see Fig. \ref{fig:drrr2-m05000-mA1250}). 
Thus CDF and DZero have an excellent chance of observing this decay.
See Fig. \ref{fig:bsmm}.
\subsubsection{Light Higgs mass}

\begin{figure}[t!]
\hspace{-.70in}
\begin{center}
\begin{minipage}{4in}
\epsfig{file=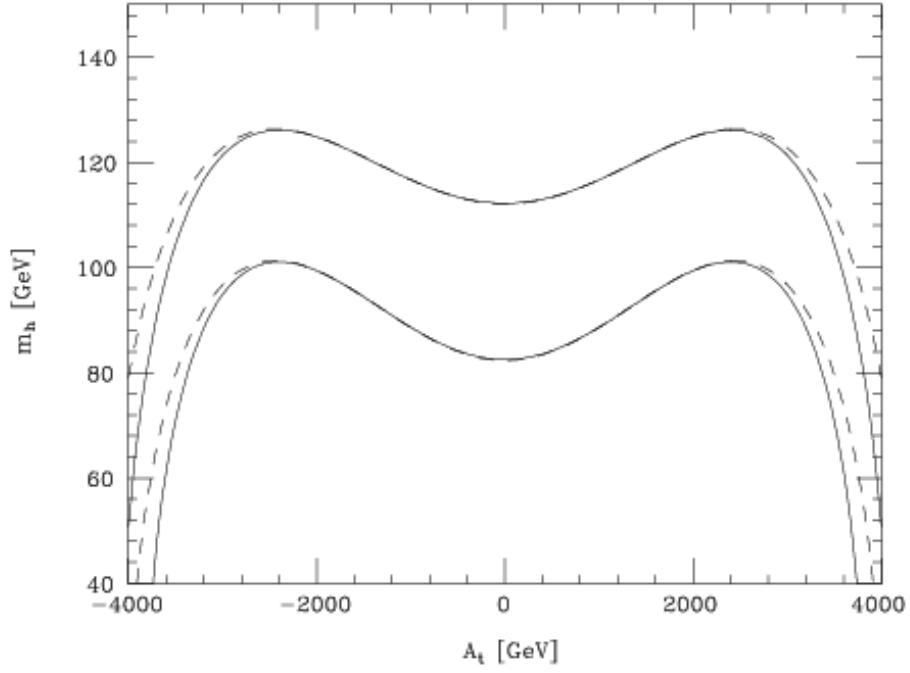,height=3.5in}
\end{minipage}
\caption{\label{fig:higgs.At} {\small The light Higgs mass as a
function of $A_t$ for two values of $\tan\beta $ = 1.5 (lower) and
15 (upper) curve \cite{Carena:1995wu}.} }
\end{center}
\end{figure}

\begin{figure}[t!]
\hspace{-.70in}
\begin{center}
\begin{minipage}{4in}
\epsfig{file=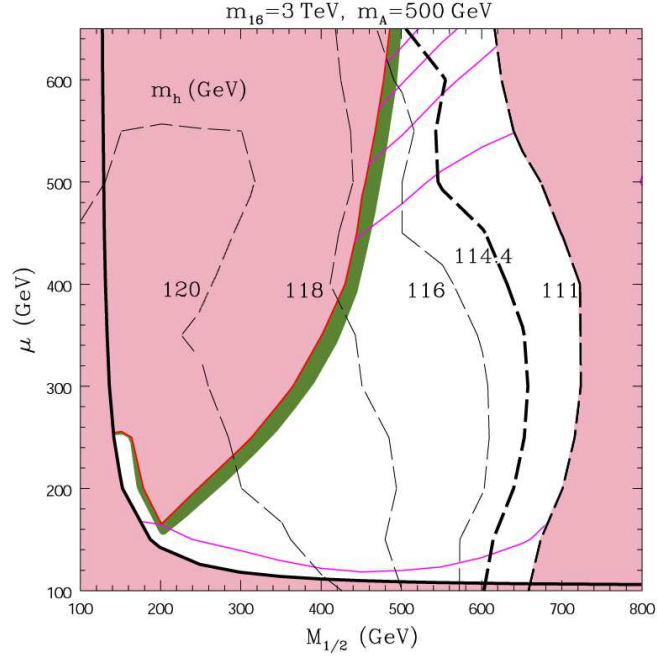,height=3.5in}
\end{minipage}
\caption{\label{fig:drrr1-m03000-mA500-mh2-wmap} {\small Light Higgs
mass contours from Ref. \cite{Dermisek:2003vn}.  Note the light
Higgs mass obtained here uses the effective field theory formalism
\cite{Carena:1995wu}.   Using FeynHiggs instead, we find values of
the light Higgs mass which are of order 3 - 6 GeV larger
\cite{Dermisek:2006dc}.} }
\end{center}
\end{figure}

It should not be obvious why the light Higgs mass decreases as
$M_{1/2}$ increases.  This is not a general CMSSM result.  In fact
this is a consequence of the global $\chi^2$ analysis and
predominantly constrained by the fit to the bottom quark mass.   The
bottom quark mass has large SUSY corrections proportional to
$\tan\beta$ \cite{yukawacorr}.  The three dominant contributions to
these SUSY corrections, $\delta m_b/m_b$ : a gluino loop
contribution $\propto \ \alpha_3 \ \mu \ M_{\tilde g} \
\tan\beta/m_{\tilde b_1}^2$, a chargino loop contribution $\propto \
\lambda_t^2 \ \mu \ A_t \ \tan\beta/m_{\tilde t_1}^2$, and a term
$\propto \ \log M_{SUSY}^2$. In addition,  in order to fit
$m_b(m_b)$  we need $\delta m_b/m_b \sim -2$ \%.   We can now
understand why the light Higgs mass decreases as $M_{1/2}$
increases.   The gluino and log contribution to the bottom mass
correction is positive for $\mu M_{\tilde g}$ positive\footnote{With
universal gaugino masses at $M_G$, both $B \rightarrow X_s \ \gamma$
and $(g-2)_\mu$ prefer $\mu \ M_{\tilde g} > 0$.}, while the
chargino loop contribution is negative, since $A_t \sim - M_{\tilde
g}$ due to an infra-red fixed point in the RG equations. And the sum
must be approximately zero (slightly negative).  Now, as $M_{1/2}$
increases, the gluino contribution also increases. Then $|A_t|$ must
also increase. However, the light Higgs mass decreases as $|A_t|$
increases (Figs. \ref{fig:higgs.At} \&
\ref{fig:drrr1-m03000-mA500-mh2-wmap}). For more details, see
\cite{Dermisek:2003vn}. In a recent paper we find the light Higgs
with mass of order $120 \pm 7$ GeV \cite{Dermisek:2006dc}.

\subsubsection{$B_u \rightarrow \tau \bar \nu$}

Recently Belle \cite{Ikado:2006un} measured the branching ratio
$Br(B_u \rightarrow \tau \bar \nu) = 1.06
\stackrel{+0.34}{-0.28}(stat) \stackrel{+0.18}{-0.16}(syst) \times
10^{-4}$.   The central value is smaller than the expected standard
model rate with $R_{B \tau \nu} \equiv \frac{Br^{exp}(B_u
\rightarrow \tau \bar \nu)}{Br^{SM}(B_u \rightarrow \tau \bar \nu)}
= 0.67 \stackrel{+0.24}{-0.21}_{exp} \pm 0.14_{|f_B|} \pm
0.10_{|V_{ub}|}$. Or the central value is $\sim 1 \ \sigma$ below
the standard model value. This may not be significant,  nevertheless
this is what is expected in the large $\tan\beta$ limit of the MSSM.
As discussed by Isidori and Paradisi \cite{Isidori:2006pk} (see
also, \cite{hou}) in the MSSM at large $\tan\beta$ one finds $R_{B
\tau \nu} \equiv \frac{Br^{SUSY}(B_u \rightarrow \tau \bar
\nu)}{Br^{SM}(B_u \rightarrow \tau \bar \nu)} = \left[ 1 -
(\frac{m_B^2}{m_{H^\pm}^2}) \frac{\tan^2\beta}{(1 + \epsilon_0
\tan\beta)} \right]$ where $\epsilon_0$ is the result of gluino
exchange at one loop.   With values of  $30 < \tan\beta < 50$, $0.5
< m_H/{\rm TeV} < 1$ and $ \epsilon \sim 10^{-2}$, they find $R_{B
\tau \nu} \equiv \frac{Br^{exp}(B_u \rightarrow \tau \bar
\nu)}{Br^{SM}(B_u \rightarrow \tau \bar \nu)} =  0.67
\stackrel{+0.30}{-0.27}$, consistent with the data.

\subsubsection{$\Delta M_{B_s}$}

CDF has recently measured the mass splitting $\Delta M_{B_s} = 17.35
\pm 0.25$ ps$^{-1}$ \cite{Abulencia:2006mq}.   Once again the
central value is below the standard model expectation $\Delta
M_{B_s} = 21.5 \pm 2.6$ ps$^{-1}$ or  $R^{exp}_{\Delta M_s} \equiv
\frac{(\Delta M_{B_s})^{exp}}{(\Delta M_{B_s})^{SM}} = 0.80 \pm
0.12$. In SUSY at large $\tan\beta$ one expects
\cite{buras,Isidori:2006pk} $R_{\Delta M_s} \equiv \frac{(\Delta
M_{B_s})^{SUSY}}{(\Delta M_{B_s})^{SM}} = 1 - m_b(\mu_b^2)
m_s(\mu_b^2) \frac{64 \pi \sin^2\theta_W}{\alpha_{em} M_A^2
S_0(m_t^2/M_W^2)} \times \frac{(\epsilon_Y \lambda_t^2
\tan^2\beta)^2}{[1 + (\epsilon_0 + \epsilon_Y \lambda_t^2)
\tan\beta]^2 [1 + \epsilon \tan\beta]^2} $. Hence  $\Delta M_{B_s}$
is suppressed at large $\tan\beta$.

\subsubsection{$B \rightarrow X_s \gamma$ and $B \rightarrow X_s
\ l^+ \ l^-$}

The effective Lagrangian for the process $B \rightarrow X_s \gamma$
is given by  $- L^{eff} \sim C_7 \ O_7$ where $O_7 = \frac{e}{16
\pi^2} m_b (\bar s_L \sigma^{\mu \nu} b_R) F_{\mu \nu}$. In SUSY  we
have $C_7 =  C_7^{SM} + C_7^{SUSY} \approx \pm C_7^{SM}$ where the
second equality is an experimental constraint, since the standard
model result fits the data to a reasonable approximation. However at
large $\tan\beta$ it was shown that the negative sign is preferred
\cite{Blazek:1997cs}.

It is then interesting that the absolute sign of $C_7$ is observable
in the process $B \rightarrow X_s \ l^+ \ l^-$.    The effective
Lagrangian for this process is given by  $- L^{eff} \sim C_7 O_7 +
C_{9, 10} O_{9,10}$ where the latter two operators are
electromagnetic penguin diagrams.     Gambino et al.
\cite{Gambino:2004mv}, using data from Belle and BaBar for $Br(B
\rightarrow X_s \ l^+ \ l^-)$, find $C_7 = + C_7^{SM}$ is preferred.
On the other hand, the Belle collaboration \cite{Abe:2005km}
analyzed the forward-backward asymmetry for the process $B
\rightarrow K^* \ l^+ \ l^-$ and find that either sign is
acceptable.   Clearly this is an important test which must await
further data.

\subsection{MSO$_{10}$SM and Large $\tan\beta$}

We conclude that the MSO$_{10}$SM
\begin{itemize}
\item fits  WMAP;
\item predicts  light Higgs with mass less than 127 GeV;
\item  predicts lighter  3rd  and heavy 1st \& 2nd  gen.
     squarks and sleptons, i.e an inverted scalar mass hierarchy;
\item  enhances the branching ratio  $Br(B_s \rightarrow \mu^+
\mu^-)$;
\item suppresses the branching ratio  $Br( B_u  \rightarrow \tau \bar \nu )$ and the mass splitting
$\Delta M_{B_s}$;
\item and favors the $sign{C_7} = - C_7^{SM}$ for the process $B \rightarrow X_s  \gamma$ which
is observable in the decay $B \rightarrow X_s \ l^+ \ l^-$.
\end{itemize}

Clearly the  MSO$_{10}$SM is a beautiful symmetry which has many
experimental tests !!

\section{Conclusion}

Supersymmetric GUTs (defined in 4,5,6 or 10 dimensions) provide a
viable, testable and natural extension of the Standard Model.  They
can also be embedded into string theory,  thus defining an
ultra-violet completion of such higher dimensional orbifold GUT
field theories. Note, that the predictions for nucleon decay are
sensitive to whether the theory is defined in four or higher
dimensions. Moreover, we showed that the minimal SO(10) SUSY model
has a wealth of predictions for low energy experiments.  Finally, I
would like to thank the organizers for this wonderful workshop.

\end{document}